# Layer Decomposition Learning Based on Gaussian Convolution Model and Residual Deblurring for Inverse Halftoning

Chang-Hwan Son

*Abstract*—Layer decomposition to separate an input image into base and detail layers has been steadily used for image restoration. Existing residual networks based on an additive model require residual layers with a small output range for fast convergence and visual quality improvement. However, in inverse halftoning, homogenous dot patterns hinder a small output range from the residual layers. Therefore, a new layer decomposition network based on the Gaussian convolution model (GCM) and structure-aware deblurring strategy is presented to achieve residual learning for both the base and detail layers. For the base layer, a new GCM-based residual subnetwork is presented. The GCM utilizes a statistical distribution, in which the image difference between a blurred continuous-tone image and a blurred halftoned image with a Gaussian filter can result in a narrow output range. Subsequently, the GCM-based residual subnetwork uses a Gaussian-filtered halftoned image as input and outputs the image difference as residual, thereby generating the base layer, i.e., the Gaussian-blurred continuous-tone image. For the detail layer, a new structure-aware residual deblurring subnetwork (SARDS) is presented. To remove the Gaussian blurring of the base layer, the SARDS uses the predicted base layer as input and outputs the deblurred version. To more effectively restore image structures such as lines and texts, a new image structure map predictor is incorporated into the deblurring network to induce structure-adaptive learning. This paper provides a method to realize the residual learning of both the base and detail layers based on the GCM and SARDS. In addition, it is verified that the proposed method surpasses state-of-the-art methods based on U-Net, direct deblurring networks, and progressively residual networks.

## I. Introduction

Printers and copiers are bilevel output devices that reproduce images on a paper by generating homogenous dot patterns using inks or toners. The printed images are in fact bilevel; however, the human visual system with the characteristics of low-pass filtering allows the printed image to be perceived as a continuous-tone image. Digital halftoning is necessitated to create a halftoned image with uniform dot patterns from a continuous-tone image with discrete gray levels (e.g., 255 gray levels) [1]. The halftoned image determines the spatial position of the inks to be deposited on a paper or controls a laser beam to form a latent image on a photoconductor drum. Digital halftoning has been used in many applications, including animated GIF generation from videos [2], removal of contour artifacts in displays [3], video processing in electronic papers [4], and data hiding [5]. The typically used digital halftoning techniques are dithering, error diffusion, and direct binary search [6].

In inverse halftoning, a continuous-tone image with 255 gray levels or more is reconstructed from its halftoned version [7]. In other words, inverse halftoning is the reverse of digital halftoning. Inverse halftoning is required in several practical applications such as bi-level data compression [8], watermarking [9,10], digital reconstruction of color comics [11], and high dynamic range imaging [12]. Inverse halftoning is an ill-posed problem with many possible solutions because digital halftoning is a many-to-one mapping. Many studies have been conducted over the last several decades, and various approaches have been introduced based on look-up tables [13], adaptive lowpass filtering [14], maximum-a-posterior estimation [15], local polynomial approximation and intersection of confidence intervals [16], and deconvolution [17]. Recently, machine-learning approaches have been actively considered based on dictionary learning [18–20] and deep convolutional neural networks (DCNNs) [21–25].

### A. Image decomposition in deep learning frameworks

Image decomposition, which is also known as layer separation in other fields, has been steadily used for image restoration [26], image enhancement [27], and image fusion [28]. Image decomposition is an approach for separating an input image into two or more layers with different gradients and illumination characteristics. Traditional image decomposition has been realized based on image transformations (e.g., wavelets) [29] and image pyramid [30] to achieve multiple resolutions. In addition, sparse representation [31], the Gaussian mixture model [32], and adaptive filtering such as bilateral [33] and guided-image filtering [34] have been used for two-layer separation, i.e., base and detail layers. In this study, the base layer corresponds to a layer whose brightness changes smoothly, resembling a low-pass-filtered image,

This work was supported by the National Research Foundation of Korea(NRF) grant funded by the Korea government(MSIT) (No. 2020R1A2C1010405).

C.-H. Son is with the Department of Software Convergence Engineering, Kunsan National University, Gunsan-si, South Korea (e-mail: changhwan76.son@gmail.com;cson@kunsan.ac.kr).

whereas the detail layer refers to a high-pass filtered image whose brightness changes rapidly. The definition of the base and detail layers may vary based on the application field.

Recently, image decomposition approaches have been incorporated into deep learning frameworks. U-net [35], Laplacian-net [36], residual networks (RNs) [37,38], and progressive residual networks (PRNs) [23,25] are representative deep learning models that apply the concept of image decomposition. U-net and Laplacian-net primarily aim to realize multiple resolutions, whereas RNs and PRNs focus on predicting residual layers. In particular, the key factor for improving image quality and accelerating convergence in an RN is that the brightness range of the residual layer should be narrow. In other words, by narrowing the output range in which the solution exists, RNs can obtain the optimal solution more easily. Therefore, it is critical to design a residual layer with a narrow brightness range.

*B. Residual layer design for residual learning*

In an end-to-end manner, RNs and PRNs are learned to map an input image into a residual layer with a narrow output range. For image restoration, the difference image between the original and input images is considered as a residual layer. Residual learning is formulated as follows:

$$x^{(r)} = f_\theta^{RN}(x^i) \approx x^o - x^i \quad (1)$$

where $x^i$, $x^o$, and $x^{(r)}$ denote the input image, original image, and predicted residual layer, respectively. Herein, parentheses in superscripts indicate predicted values. Bold and italic lowercase letters indicate vectors. $f_\theta^{RN}$ indicates the DCNN with parameter $\theta$ for estimating the residual layer. As shown in Eq. (1), the output of the network is the residual, and it differs from those of conventional DCNNs that directly transform the input image $x^i$ to the original image $x^o$ with a relatively wide output range. In addition, the residual layer is designed as a difference image between $x^o$ and $x^i$, as shown in Eq. (1). This is because the measured input images can be modeled physically as the addition of original images and residual layers.

$$x^i = x^o + x^r \quad (2)$$

where $x^i$ indicates the measured input images. For example, captured noisy images and rain images can be measured images. $x^r$ is the residual layer that contains artifacts such as noise and rain streaks. The residual layer $x^o - x^i$, as shown in Eq. (1), is derived from the additive model of Eq. (2).

Previous studies [37,38] showed that using the difference image as a residual layer can effectively improve visual quality and increase convergence speed. For example, in image denoising, the noise layer is used as the residual layer, which corresponds to the difference image between the original and noisy images. In general, noise is assumed to exhibit a Gaussian distribution. This implies that most of the pixels in the noisy layer are zero. Therefore, the output range of the noise layer can be narrowed. In rain removal, the rain layer including only rain streaks is used as the residual layer, and it is obtained by subtracting the original image from the input rain image. Because the rain layer includes only rain streaks, a narrow output range can be guaranteed in the residual layer.

*C. Residual learning problems for inverse halftoning*

Digital halftoning is a nonlinear system that includes binary quantization. Therefore, the additive model, as shown in Eq. (2), is no longer valid for digital halftoning, that is, $x^i \neq x^o + x^r$. This means that residual learning, as shown in Eq. (1), cannot be directly applied to inverse halftoning. More specifically, the halftoned image is a bilevel image composed of black and white dot patterns. If the residual layer is defined as the difference image between the original image and the input halftoned image, similar patterns that appear as black and white dot patterns can appear in the residual layer. Inevitably, a sudden change in brightness is accompanied by a residual layer. Hence, merely creating a residual layer based on image difference is not suitable for inverse halftoning.

*D. Progressively residual learning problems for inverse halftoning*

Progressively residual learning (PRL) [23,25] can be an alternative for solving sudden changes in brightness, as mentioned in the previous subsection. In PRL, the base layer whose brightness changes smoothly is first recovered; subsequently, the remaining detail layer is predicted.

$$x^{(d)} = f_\theta^{PRL\_r}(x^{(b)}, x^{(i)}) \approx x^o - x^{(b)},$$
$$\text{where } x^{(b)} = f_\theta^{PRL\_b}(x^i) \quad (3)$$

$x^{(b)}$ and $x^{(d)}$ indicate the predicted base layer and detail layer, respectively. For inverse halftoning, the input halftoned image $x^i$ cannot be used as the base layer. However, in PRL, the input halftoned image $x^i$ is first converted into the base layer $x^{(b)}$ through the pretrained DCNN $f_\theta^{PRL\_b}$. The generated base layer resembles a lowpass-filtered image, and it can be considered as an approximation of the original image. If the detail layer, which is defined as $x^o - x^{(b)}$, is used as the residual layer, then a narrow brightness range can be guaranteed. This implies that residual learning, $f_\theta^{PRL\_r}$, is possible. The additive model of Eq. (2) can be used reasonably with PRL for inverse halftoning. For reference, the input halftoned image $x^{(i)}$ can be used with the base layer $x^{(b)}$, as shown in Eq. (3), to estimate the detail layer, thereby compensating for information loss in the predicted base layer.

However, PRL [23,25] applied to inverse halftoning has not been able to present a new deep learning model from the viewpoint of creating base and detail layers. In PRL, $f_\theta^{PRL\_b}$ is trained to generate the base layer. However, the output images of $f_\theta^{PRL\_b}$ cannot be regarded as the base layer. Instead, the output images correspond to the final reconstructed images because they appear similar to the original images. Moreover, the predicted base layers appear better visually than the reconstructed images using traditional inverse halftoning methods based on dictionary learning [19] and look-up tables [13]. If the image quality of the base layers decreases to the level of Gaussian blurring of the original images, then conventional PRL cannot yield satisfactory results. In summary, the PRL developed for inverse halftoning hitherto merely applies inverse halftoning twice in succession.

*E. Contributions*

- This paper presents three major points. In particular, a new method for creating base and detail layers based on the proposed structure-aware layer decomposition learning (SALDL) is introduced. First, to design the base layer, a new statistical distribution of image difference between a blurred continuous-tone image and a blurred halftoned image with a Gaussian filter with a narrow output range is shown. Based on this observation, the base layer is reconstructed using a new GCM-based residual subnetwork that predicts the difference between the blurred continuous-tone image and blurred halftoned image; this method differs completely from the existing PRL [23,25], which uses an initial restored image from a DCNN for base layer generation.
- Second, the detail layer is generated based on structure-aware residual learning that predicts the difference image between the predicted base layer and the original image. To more effectively enhance image structures such as edges and textures, an image structure map predictor, which has been introduced in a previous study [24], is incorporated into the residual detail layer learning, thereby resulting in structure-enhancing learning. In addition, the predicted base layer is the lowpass-filtered version of the original image. Therefore, the proposed residual detail learning should be learned to deblur the base layer, i.e., to remove the blurring of the base layer. This implies that the deblurring strategy is adopted in the proposed residual detail learning, unlike the existing PRL.
- Third, it is demonstrated through SALDL can recover high-quality images from the predicted base layers, whose quality is poor in terms of edge and texture representation. However, the existing PRL [23,25] cannot yield satisfactory results from the same base layers. This reveals that the existing PRL is not suitable for low-quality base layers. By contrast, the proposed structure-aware residual learning method is more effective for describing image structures. To our best knowledge, this is the first study that performed the abovementioned comparison, and the experimental results confirmed the feasibility of the proposed SALDL as a new PRL for inverse halftoning that surpasses state-of-the-art methods such as PRL, U-net, and DCNN.

## II. PROPOSED SALDL BASED ON GCM

*A. Motivations*

Image decomposition is an approach for analyzing and reconstructing images. Image transformation (e.g., wavelet transformation), structure-adaptive filtering, and sparse coding have been considered as effective tools for realizing image decomposition. However, DCNNs have recently demonstrated excellent performances in image enhancement and restoration.

Therefore, this study focuses on incorporating image decomposition into a deep learning framework for inverse halftoning. In particular, a new deep learning model to enable the residual learning of both the base and detail layers is introduced. As discussed in the Introduction, residual learning that directly maps an input image into the residual layer is not applicable to inverse halftoning because the additive model is no longer valid. Moreover, the output range of the residual layer cannot be narrowed owing to black and white dot patterns. PRL can be considered as an alternative for realizing image decomposition. However, the PRL that has been developed for inverse halftoning hitherto merely applies inverse halftoning twice in succession, since the quality level of the restored base layer is similar to that of the original image. In addition, the PRL merely uses initially reconstructed images through a DCNN for base layer generation; hence, the design of the base layer lacks novelty. Furthermore, existing PRL cannot recover textures and fine details from low-quality base layers. Hence, a new SALDL based on the GCM is proposed herein.

Fig. 1 shows the concept of image decomposition based on the proposed SALDL for inverse halftoning. Unlike traditional approaches such as wavelet transform and image pyramid, residual-learning-based image decomposition is proposed. In particular, novel GCM-based residual learning and structure-aware residual deblurring are introduced for base and detail layer generation, respectively. By adding the predicted base and detail layers, a continuous-tone image can be reconstructed from the input halftoned image. Details regarding the generation of the base and detail layers are provided below.

*B. Residual layer design for baser layer generation*

Unlike the residual layer design based on the additive model of Eq. (1), a new GCM is proposed herein to generate the residual of the base layer.

$$x^{r_b} = x^o \otimes k^g - x^i \otimes k^g = (x^o - x^i) \otimes k^g \quad (4)$$

where $x^{r_b}$ denotes the residual layer corresponding to the base layer. Herein, the base layer is defined as the Gaussian blurring of the input halftoned image, $x^i \otimes k^g$. Here, $\otimes$ denotes the convolution operation, and $k^g$ indicates the Gaussian smoothing filter. Therefore, Eq. (4) indicates that the residual layer corresponding to the base layer is defined as the image difference between the blurred original image and blurred halftoned image through Gaussian filtering. Compared with Eq. (1), the proposed residual layer is the filtered version of $x^o - x^i$. Hereinafter, the proposed model expressed as Eq. (4) is referred to as GCM to differentiate it from the additive model expressed in Eq. (1).

The main objective of residual learning is to narrow the output range. Whether the residual layer generated based on the GCM yields a narrow output range is yet to be elucidated. The histogram distribution for one sample image was analyzed to verify this. Fig. 2 shows four images for generating two types of residual layers. The original, halftoned, blurred original, and blurred halftoned images are shown from left to right. Fig. 3 shows a comparison of the histogram distributions for the two types of residual layers. One is the residual layer generated using the additive model, which subtracts the original image from the halftoned image. The other is the residual layer generated using the proposed GCM, which subtracts the blurred original image from the blurred halftoned image. As shown in the histogram distributions, the residual layer generated using the proposed GCM yielded a narrow output range compared with the conventional additive model, which yielded a wider output range. This is because the residual layer generated based on the additive model tends to exhibit textures that resemble dot patterns. Meanwhile, the proposed GCM utilizes Gaussian filtering to smooth out sudden changes that appear in halftoned images, thereby enabling the output range of the residual layer to be narrow.

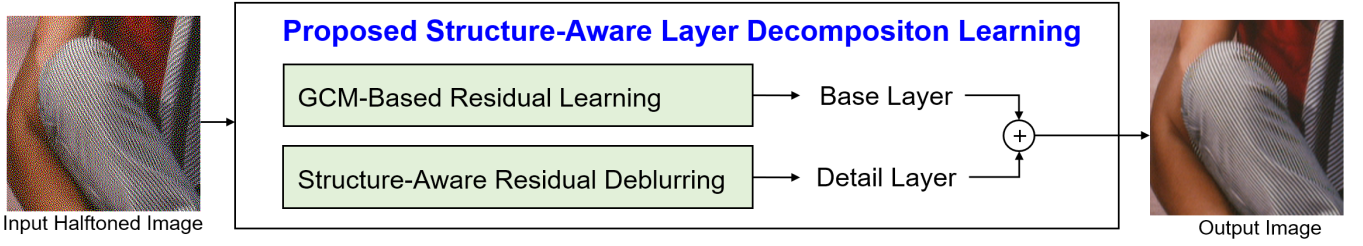

Fig. 1. Concept of image decomposition based on proposed SALDL for inverse halftoning.

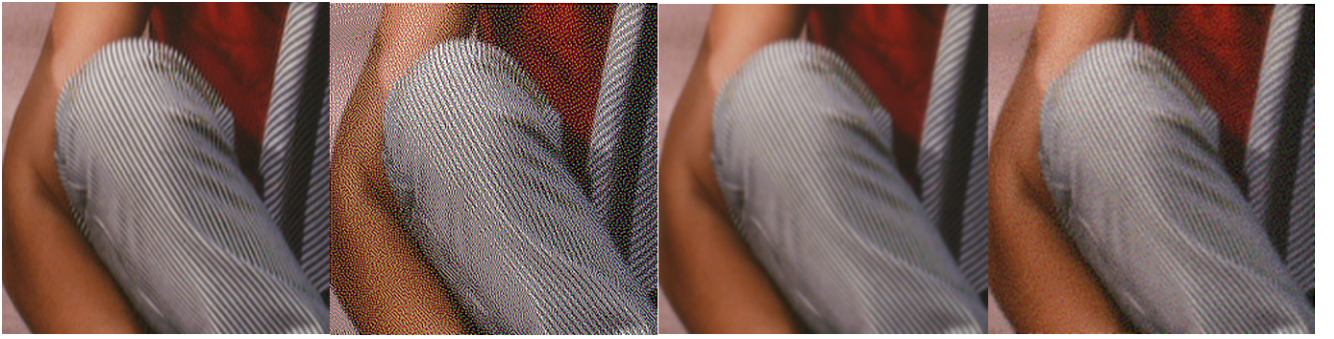

Fig. 2. Original, halftoned, blurred original, and blurred halftoned images (left to right).

### C. GCM-based residual subnetwork for baser layer generation

To realize the proposed GCM for base layer generation, a GCM-based residual subnetwork was designed, as shown in Fig. 4. To implement the proposed GCM, as shown in Eq. (4), Gaussian filtering was first applied to the input halftoned image. In existing deep learning tools, it can be easily implemented through a convolution layer where the convolution filter is fixed as a Gaussian filter. The Gaussian-filtered halftoned image was passed through the GCM-based residual subnetwork to output the residual layer.

$$x^{(r_b)} = f_\theta^{GCM}(x^i \otimes k^g) \qquad (5)$$

where $x^{(r_b)}$ is the predicted residual layer for base layer generation, and $f_\theta^{GCM}$ denotes the GCM-based residual subnetwork to be trained. Herein, parentheses in superscripts indicate the predicted values. The standard deviation of the Gaussian filter $k^g$ was set to 1 and the filter size was $5 \times 5$.

To train $f_\theta^{GCM}$, a loss function is defined as follows:

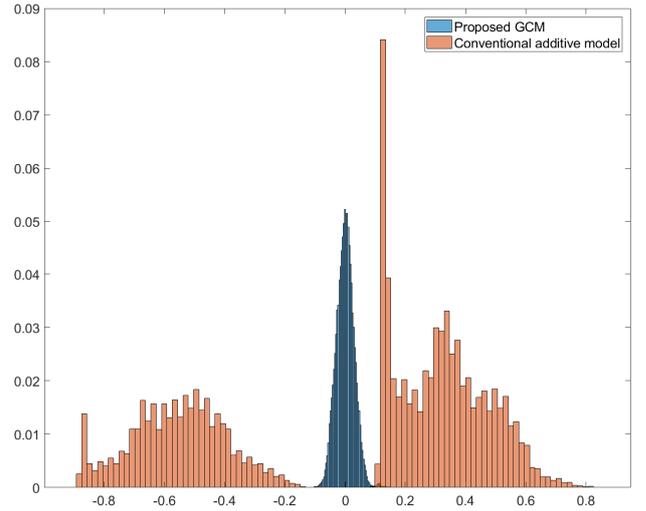

Fig. 3. Histogram comparison of two types of residual layers generated using conventional additive model and proposed GCM.

$$L = \frac{1}{m}\sum_{i=1}^{M} \left\| x_i^{(r_b)} - x_i^{r_b} \right\|^2 \qquad (6)$$

where $i$ denotes a training sample, $M$ is the batch size, and $\|\cdot\|$ is the $l_2$-norm. Compared with the additive model, the proposed GCM-based residual subnetwork can narrow the output range of the residual layer.

For the pretrained GCM-based residual subnetwork, the base layer was generated as follows:

$$x^{(b)} = x^{(r_b)} + x^i \otimes k^g \qquad (7)$$

where $x^{(r_b)}$ is the output of the pretrained GCM-based residual subnetwork $f_\theta^{GCM}$, and $x^{(b)}$ is the predicted base layer. This equation indicates that the base layer is the addition of the Gaussian-filtered halftoned image and the predicted residual layer through the GCM-based residual subnetwork. For reference, the entire architecture, as shown in Fig. 3, was not trained. Based on heuristic experiments, it was discovered that the learning of the entire architecture did not yield good results.

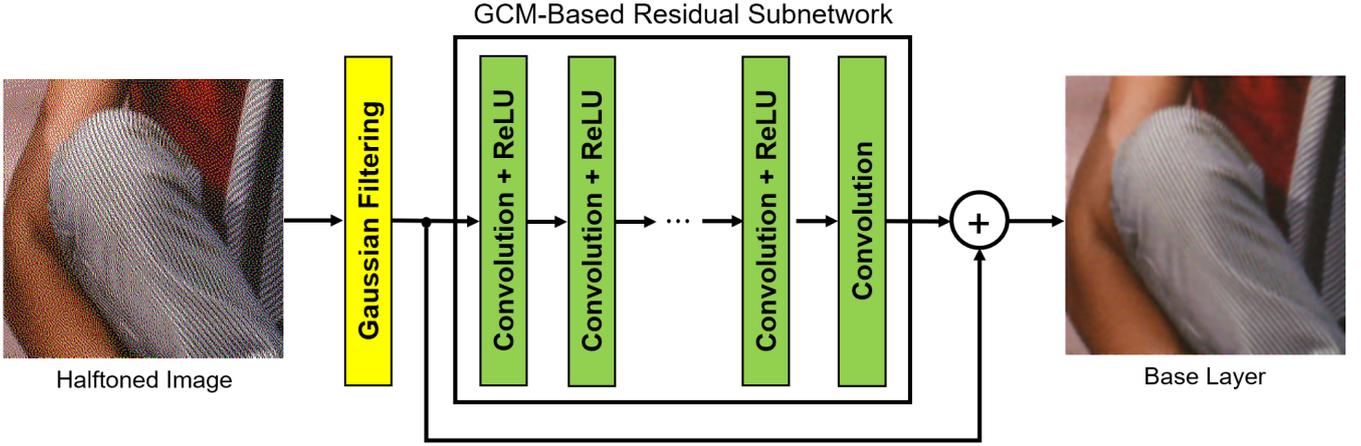

Fig. 4. GCM-based residual subnetwork for base layer generation.

*D. Detail layer design*

The predicted base layer is the approximation of the Gaussian-filtered original image.

$$x^{(b)} \approx x^o \otimes k^g \quad (8)$$

As shown in Fig. 4, details such as textures and edges were absent in the predicted base layer; however, it contained the low-frequency components of the original image. Therefore, the detail layer to be predicted was designed based on the difference between the original image and the predicted base layer.

$$x^d = x^o - x^o \otimes k^g \approx x^o - x^{(b)} \quad (9)$$

The predicted base layer $x^{(b)}$ was regarded as an approximation of the Gaussian-filtered original image $x^o$. This implies that the detail layer $x^d$ contains textures and edges with small pixel values, and hence the brightness range of the detail layer is narrow. According to the detailed layer design based on the proposed GCM, residual learning can be performed for the detail layer.

*E. Direct deblurring approach*

The predicted base layer is the approximation of the Gaussian-filtered original image, as shown in Eq. (8). Therefore, conventional image deblurring methods can be considered to directly reconstruct the original image from the predicted base layer. Conventional image deblurring methods can restore the missing details by removing the Gaussian blurring of the predicted base layer. Image deblurring problems [28] can be formulated as follows:

$$arg\min_{x} \|x^{(b)} - x^o \otimes k^g\|^2 + \lambda \sum_{j=1}^{2} \|x^o \otimes k^{h,j}\|^\alpha \quad (10)$$

where $k^{h,j}$ indicates high-pass filters such as horizontal and vertical filters. $\alpha$ controls the sparsity, and $\lambda$ is a constant to weight the regularization term [28]. In general, the motion kernel $k^g$ in Eq. (10) is unknown; however, a Gaussian filter $k^g$ can be used for the motion kernel based on the proposed GCM. Additionally, the motion kernel can be estimated directly from the base layer. This case corresponds to blind image deblurring. It appears that conventional image deblurring can yield good results. However, some issues exist. A comparison between Figs. 2 and 4 shows that the predicted base layer differs from the blurred original image. In other words, textures and edges are missing, and noise is generated. In addition, the noises differed from the Gaussian random noise, which has been considered to solve the image deblurring problem. Therefore, conventional image deblurring methods are not suitable for restoring the original image from the predicted base layer.

In another image deblurring approach, deep learning tools are used. More specifically, the DCNN can be trained to transform the predicted base layer to the original continuous-tone image [39].

$$x^{(o)} = f_\theta^{DDN}(x^{(b)}) \quad (11)$$

where $f_\theta^{DDN}$ denotes the direct deblurring network (DDN), and $x^{(o)}$ is the reconstructed continuous-tone image. Because the predicted base layer $x^{(b)}$ is the Gaussian-blurred version of the original image, $f_\theta^{DDN}$ is regarded as a deblurring network. Because the predicted base layer has already lost some textures and sharpness, the input halftoned image $x^i$ can be used as additional information.

*F. Proposed layer decomposition learning*

In addition to the DDN, as shown in Eq. (11), the residual deblurring strategy can be adopted. It is noteworthy that the DDN and residual deblurring network (RDN) were derived from the proposed GCM. In other words, both are the proposed deep-learning architectures. The RDN estimates the detail layer from two types of images, i.e., the input halftoned image and the predicted base layer via residual learning. It appears that the RDN is similar to the conventional PRL[23,25]. However, the significant difference is that the deblurring strategy is adopted in the former. In other words, the predicted base layer is the Gaussian-filtered version of the original image, and the base layer is designed based on the proposed GCM for residual learning. This RDN can provide better performances than the DDN, owing to the effect of residual learning. However, this RDN is restricted in terms of recovering image structures clearly. Hence, a new subnetwork known as the image structure map predictor is incorporated in the proposed SALDL.

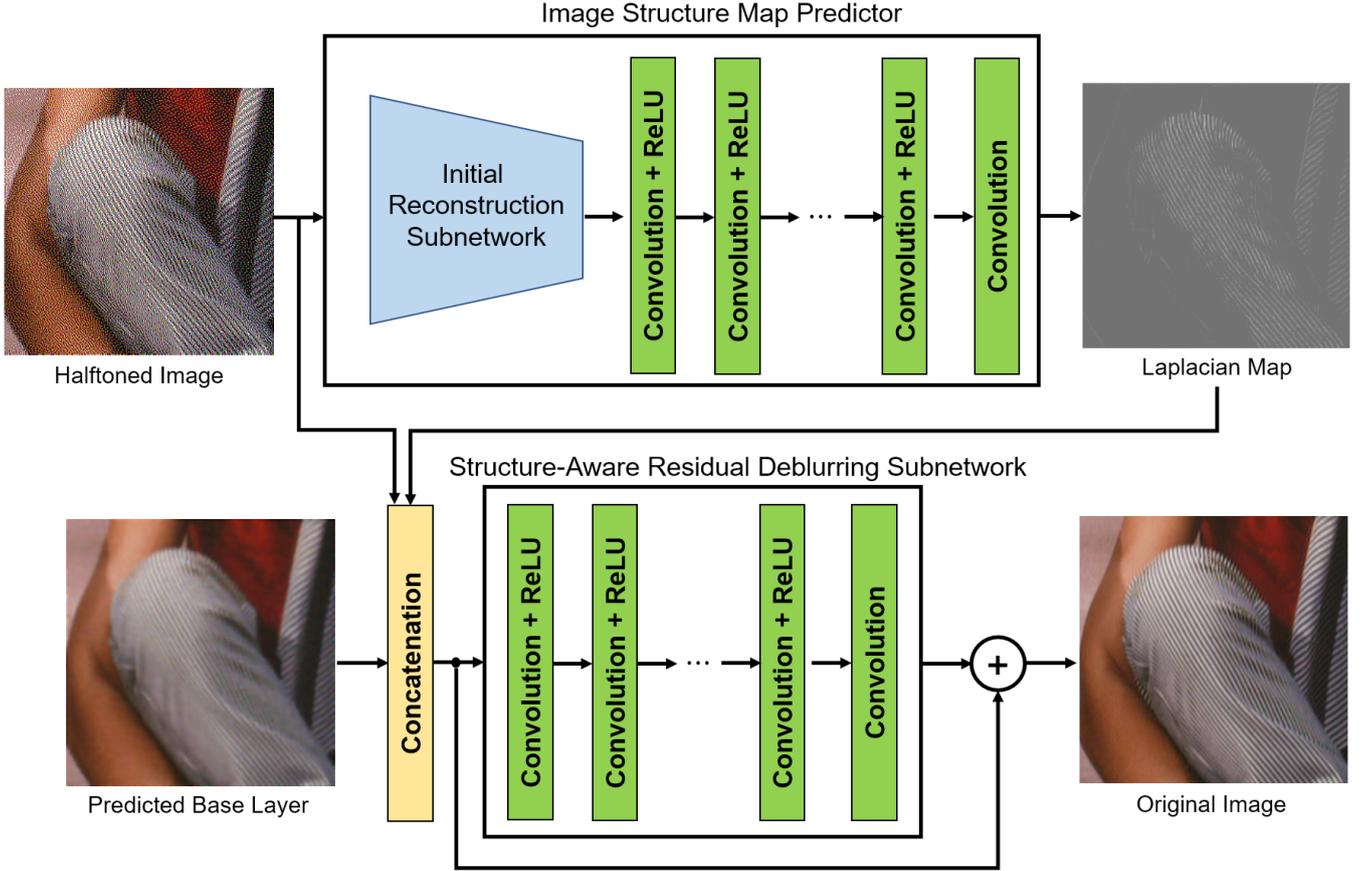

Fig. 5. Proposed SALDL for inverse halftoning.

Fig. 5 shows the entire architecture of the proposed SALDL, which comprises two subnetworks. One is the image structure map predictor (ISMP), and the other is the SARDS. The ISMP transforms the input halftoned image into a Laplacian map, which refers to an image obtained by convolving the original image and the Laplacian filter. An example of the predicted Laplacian map is shown on the right side of Fig. 5. Even though the predicted base layer can be input to the image structure map predictor, in this case, the detailed representation is not satisfactorily restored because the predicted base layer has already lost some texture information. As shown in Fig. 5, the input halftoned image contains more texture information than the predicted base layer.

The ISMP includes a pretrained subnetwork known as the initial reconstruction subnetwork (IRS). This subnetwork generates the initial reconstructed image from the input halftoned image. Because the input halftoned image is quantized, it is preferable to predict the image structures from the initial reconstructed image than from the halftoned image. In fact, the Laplacian map is the filtered version of the original image, which implies that the Laplacian map can be predicted by convolving the Laplacian filter with the initial reconstructed image. However, the initial reconstructed image differs from the original image; hence, more convolution and ReLU layers are required at the back of the IRS. Based on experiments, it was confirmed that the accuracy of the Laplacian map decreased when the IRS was not adopted, rendering the predicted detail layer less accurate. Therefore, the IRS is key for increasing the accuracy of the ISMP. As shown in Fig. 5, the initial reconstructed image was changed to increase the performance of the ISMP while learning the entire network.

The SARDS requires three input images: the predicted base layer, Laplacian map, and input halftoned image. The predicted Laplacian map was stacked on the top of the input halftoned image and the predicted base layer via a concatenation layer; subsequently, it was input to the SARDS to estimate the detail layer.

$$\boldsymbol{x}^{(d)} = f_\theta^{sards}\big(\boldsymbol{x}^{(b)}, \boldsymbol{x}^{(l)}, \boldsymbol{x}^i\big), \ \boldsymbol{x}^{(l)} = f_\theta^{ismp}(\boldsymbol{x}^i) \quad (12)$$

where $f_\theta^{sards}$ and $f_\theta^{ismp}$ denote the proposed SARDS and ISMP, respectively. $\boldsymbol{x}^{(l)}$ denotes the predicted Laplacian map, and $\boldsymbol{x}^{(d)}$ denotes the predicted detail layer. In Eq. (12), the Laplacian map is predicted from the input halftoned image, not

arXiv:2012.13894 [eess.IV]

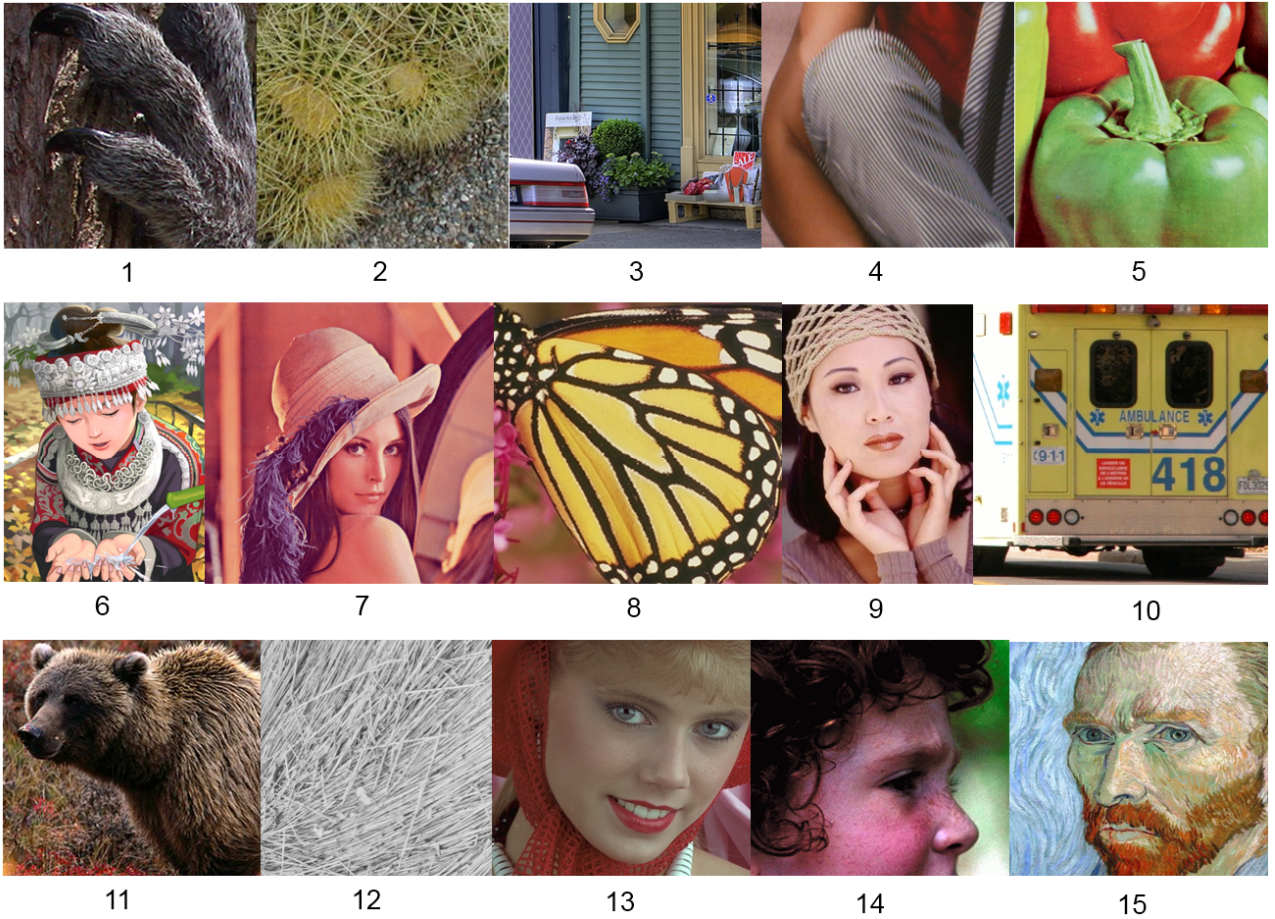

Fig. 6. Test images.

TABLE I. NUMBER OF FILTERS AND CHANNELS USED IN CONVOLUTIONAL LAYERS

| Subnetworks \ Layers | Input layer | Last convolution layer | Other layers |
|---|---|---|---|
| GCM-based residual subnetwork | $c=1, m=64$ | $c=64, m=1$ | $c=64, m=64$ |
| ISMP | $c=1, m=64$ | $c=64, m=1$ | $c=64, m=64$ |
| SARDS | $c=3, m=64$ | $c=64, m=1$ | $c=64, m=64$ |

the base layer. Based on experiments, it was discovered that the Laplacian map was not accurately estimated because the base layer contained missing information. The use of the Laplacian map provided subnetwork $f_\theta^{sards}$ with spatial information regarding areas that were flat, lined, or textured. This information enabled the entire network to be trained by adapting to local image structures. Consequently, the texture representation of the detail layer can be improved and noisy dot patterns on flat areas can be removed effectively. The ISMP can be regarded as a type of attention network, whereas the predicted Laplacian map is in fact a spatial attention feature map.

The multiloss function was used to learn $f_\theta^{sards}$ and $f_\theta^{ismp}$; it is expressed as

$$L = \frac{1}{m}\sum_{i=1}^{M} \omega_1 \|x_i^{(d)} - x_i^d\|^2 + \omega_2 \|x_i^{(l)} - x_i^l\|^2 \quad (13)$$

where $i$ denotes the training sample, $M$ the batch size of, and $\omega$ the weight of the two subnetworks. As shown in Eq. (12), the accuracy of $f_\theta^{ismp}$ affects the accuracy of $f_\theta^{sards}$. Therefore, in this study, $\omega_1$ and $\omega_2$ were set to the value of 1.

For the trained $f_\theta^{sards}$ and $f_\theta^{ismp}$, the final continuous-tone image was generated based on the additive model, i.e., $x^{(o)} = x^{(d)} + x^{(b)}$. As mentioned in the Introduction, the additive model is not suitable for inverse halftoning. However, by generating the Gaussian-blurred version of the original image, layer decomposition learning based on the GCM and SARDS can be applied to inverse halftoning.

### III. EXPERIMENTAL RESULTS

The proposed SALDL for inverse halftoning was implemented using MatConvNet [40] and trained with two 2080Ti GPUs on a Windows operating system. To evaluate the proposed method, it was compared with state-of-the-art deep learning methods based on the DCNN [37], DDN[39], U-Net [35], and PRL [23,25]. In this study, a Gaussian-blurred halftoned image was used as the base layer in both the DNN and PRL methods to implement Eqs. (11) and (3), respectively: In other words, the same base layer was used for pair comparison. This can reveal the effectiveness of the proposed method in recovering image structures compared with the DDN and PRL. For performance evaluation, the peak signal-to-noise ratio (PSNR) and structure similarity (SSIM) [41] were used to measure the inverse of the MSE in a log space and the structure similarity between two images, respectively. For both the PSNR and SSIM, a higher value indicates higher quality. The source codes of the proposed SALDL can be downloaded at https://sites.google.com/view/chson.

*A. Training data collection*

For training, public datasets [36] including General 100, Urban 100, BSDS100, and BSDS200 were used to prepare continuous-tone color images. The total number of continuous-tone color images was 500. General 100, urban 100, and BSDS200 were used for training, whereas BSDS100 was used for validation. The same training and validation sets were used to train all the deep-learning-based methods: the proposed SALDL, PRL, U-net, DDN, and DCNN. The three subnetworks of the GCM-based residual subnetwork, IRS, and SARDS used the same training and validation datasets. For digital halftoning, the continuous-tone color images were converted into grayscale images; subsequently, error diffusion [42] was used to transform the grayscale images into halftoned images. The Floyd–Steinburg filter [42,1] was used for error diffusion. The Laplacian operator was applied to the grayscale images to obtain Laplacian maps. To obtain the training patches, three types of patches were extracted randomly from the grayscale original images, Laplacian maps, and halftoned images. The extracted patch is of size 32 × 32. In this study, grayscale patches were used for training because error diffusion can be easily applied to them. To apply the proposed trained network to color images in the test phase, the color image was first separated into R, G, and B planes; subsequently, the proposed network was applied to each plane independently.

*B. Networking Training*

All the subnetworks including the GCM-based residual subnetwork, ISMP, and SARDS comprised convolution and ReLU layers. Hereinafter, a pair comprising convolution and ReLU layers is known as a convolution block. In the subnetworks, $m$ filters measuring $5 \times 5 \times c$ were used in the convolutional layers. Here, $c$ represents the number of input channels. Table I shows the number of filters and channels used in the convolutional layers. In the input layer of the RS, $c$ was set to 3 because three input channels of the base layer, the Laplacian map, and the halftoned image were input to the input layer. The filters were initialized using a random number generator. The numbers of convolution blocks used in the GCM-based residual subnetwork, IRS, and SARDS were set to 16. The number of convolution blocks used in the ISMP except the IRS was 6. To update the convolution filters, the mini-batch gradient descent algorithm [43] was used. The epoch number was 200, and the batch size was 64. Each epoch involved 1,000 backpropagation iterations. The learning rate began at $10^{-5}$ and decreased linearly every 50 epochs to $10^{-6}$. All loss functions were modeled by the $l_2$ norm.

*C. Visual quality evaluation*

Fig. 6 shows the 15 test images number accordingly for visual quality evaluation. The test images contained various types of image structures, including lines, curves, and regular patterns, to verify whether the proposed SALDL can improve the detail representation and dot elimination. Certainly, the test images were not included in the training and validation data sets. Fig. 7 shows the experimental results. As shown in the red boxes, the proposed method describes the image structures more accurately. In addition, the overall sharpness of the images was better. In particular, the lines of the pants were restored in more detail and were sharper (as shown in the first row) when using the proposed method. The second row shows more clearly expressed cactus thorns. The third row shows the textures on the palm and the hair accessory in more detail. As shown in the fourth, fifth, sixth, and seventh rows, texts including the license plate, rip outline, straw, and Gogh's eyes, respectively, were restored more clearly. Moreover, as shown in the blue box in the fifth row, the proposed method suppressed noisy dots on flat areas, unlike the case involving the conventional DCNN [37] and U-Net [35] methods. The blue box of the last row shows that the proposed method can reproduce smooth skin tones in the face areas, whereas the face areas reconstructed using other methods appeared rougher and noisier.

By comparing the proposed method with the DDN/PRL methods, it was verified that the additional use of the ISMP can improve the performance for detailed representation and dot elimination. The DDN directly predicts the continuous-tone images from the base layers, as shown in Eq. (11). Because the base layers are predicted, some information may be lost. Hence, the flat areas of the reconstructed images appeared slightly noisy, and the sharpness can be further improved. The PRL method additionally uses input halftone images to increase the amount of information for residual learning, as shown in Eq. (3). Therefore, the PRL method can provide results with improved image quality compared with the DDN method. However, the PRL method lacks image structure representation. In addition, the existing PRL cannot produce satisfactory results from the same base layers. This reveals that the architecture of the existing PRL is not suitable for low-quality base layers. Hence, the proposed SALDL uses the ISMP to identify Laplacian maps from the input halftoned images. The Laplacian map provides the SARDS with spatial information regarding areas that are flat, lined, or textured. This information enables the proposed SALDL to be adaptive to local image structures. Consequently, the texture representation of the detail layer can be improved and noisy dot patterns on flat areas can be suppressed effectively. The ISMP can be regarded as a type of attention network, and the predicted Laplacian map is a spatial attention feature map.

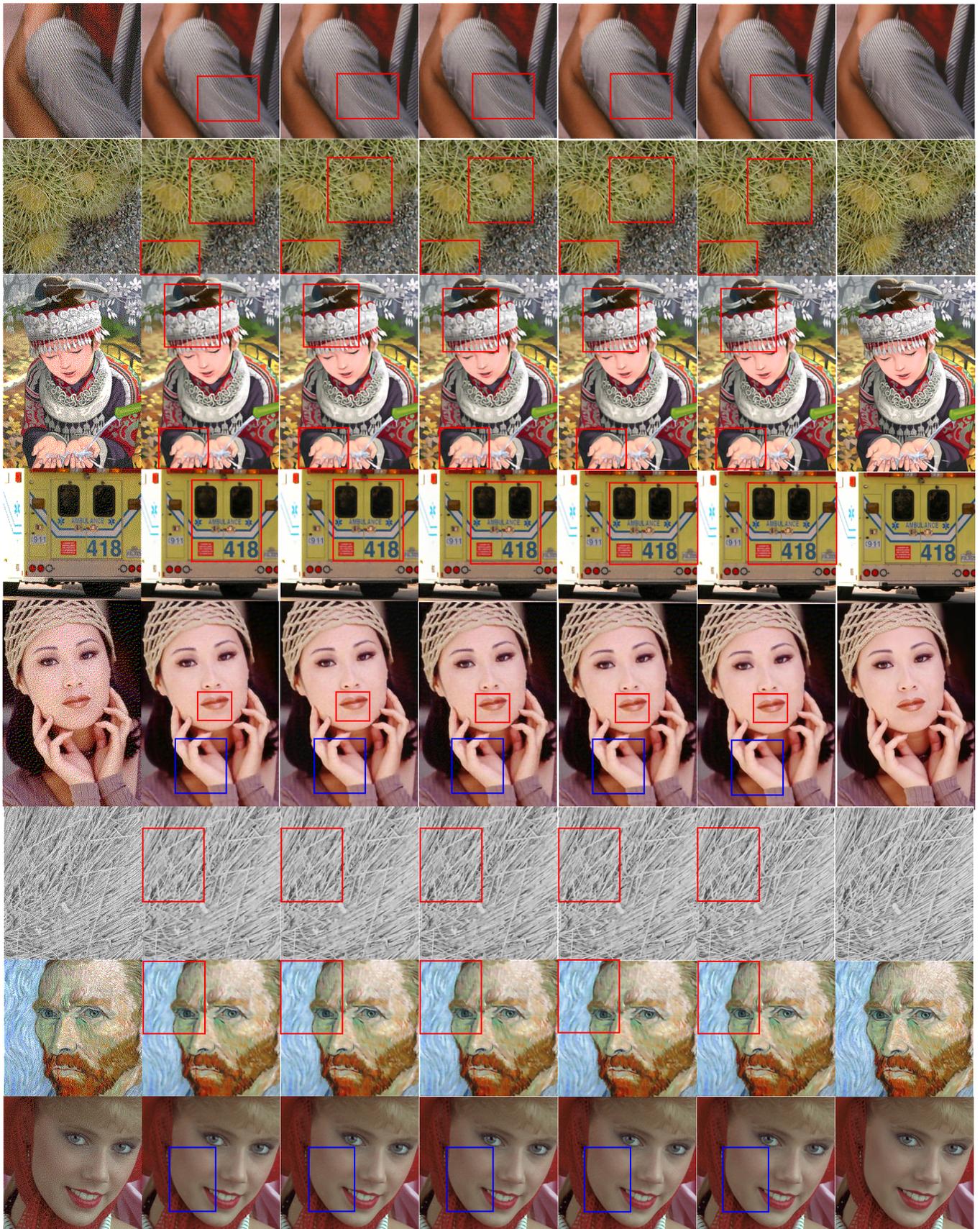

Fig. 7. Experimental results: halftoned images, images reconstructed using DCNN [37], images reconstructed using U-net [35], images reconstructed using DDN [39], images reconstructed using PRL [23,25], images reconstructed using proposed SALDL method, and original images (left to right).

TABLE II. PERFORMANCE EVALUATION.

| Methods | Proposed Method | | U-Net [35] | | DCNN [37] | | DDN [39] | | PRL [23,25] | |
|---|---|---|---|---|---|---|---|---|---|---|
| Test Images | PSNR | SSIM | PSNR | SSIM | PSNR | SSIM | PSNR | SSIM | PSNR | SSIM |
| 1 | 25.943 | 0.835 | 25.563 | 0.815 | 25.181 | 0.808 | 24.747 | 0.783 | 25.639 | 0.818 |
| 2 | 25.916 | 0.913 | 25.590 | 0.904 | 25.395 | 0.900 | 25.139 | 0.894 | 25.632 | 0.905 |
| 3 | 26.013 | 0.878 | 25.247 | 0.857 | 24.810 | 0.846 | 24.308 | 0.826 | 25.379 | 0.859 |
| 4 | 29.951 | 0.901 | 29.262 | 0.873 | 28.608 | 0.854 | 28.997 | 0.864 | 28.843 | 0.866 |
| 5 | 31.974 | 0.981 | 31.901 | 0.979 | 31.818 | 0.979 | 31.064 | 0.978 | 31.488 | 0.979 |
| 6 | 26.373 | 0.909 | 25.820 | 0.899 | 25.370 | 0.890 | 24.974 | 0.88 | 25.814 | 0.896 |
| 7 | 31.601 | 0.981 | 31.248 | 0.979 | 31.084 | 0.979 | 30.522 | 0.977 | 31.069 | 0.979 |
| 8 | 28.659 | 0.969 | 27.992 | 0.966 | 27.275 | 0.959 | 26.698 | 0.953 | 27.823 | 0.963 |
| 9 | 31.145 | 0.953 | 30.539 | 0.948 | 30.237 | 0.949 | 29.517 | 0.933 | 30.449 | 0.942 |
| 10 | 30.281 | 0.939 | 29.601 | 0.930 | 29.214 | 0.928 | 28.721 | 0.914 | 29.581 | 0.929 |
| 11 | 24.853 | 0.859 | 24.098 | 0.832 | 23.738 | 0.828 | 23.388 | 0.805 | 24.258 | 0.839 |
| 12 | 25.654 | 0.816 | 24.718 | 0.751 | 24.441 | 0.739 | 24.274 | 0.741 | 24.904 | 0.771 |
| 13 | 33.381 | 0.966 | 33.302 | 0.964 | 33.282 | 0.964 | 32.426 | 0.959 | 32.777 | 0.961 |
| 14 | 29.901 | 0.846 | 29.631 | 0.840 | 29.753 | 0.832 | 29.253 | 0.822 | 29.645 | 0.833 |
| 15 | 27.119 | 0.904 | 26.878 | 0.897 | 26.755 | 0.894 | 26.422 | 0.89 | 26.841 | 0.898 |
| AVG. | 28.584 | 0.910 | 28.093 | 0.896 | 27.797 | 0.890 | 27.363 | 0.881 | 28.009 | 0.896 |

Based on Eqs. (3) and (11), the DDN and PRL methods use the same base layers generated using the proposed GCM-based residual subnetwork. The DDN is one of the proposed deep learning architectures for inverse halftoning because it is derived from the proposed GCM to predict Gaussian-blurred images. In the existing PRL methods, no specific models exist for the residual learning of the base layer. In addition, the existing PRL cannot produce satisfactory results from low-quality base layers. To our best knowledge, this study is the first to perform the abovementioned comparison, and the experimental results confirmed that the proposed SALDL can be used as a new deep learning model for inverse halftoning that enables residual learning for both the base and detail layers by incorporating image decomposition into the deep learning framework.

Table II shows the results of the PSNR and SSIM evaluations. As expected, the proposed SALDL method demonstrated the best performance among all the methods and surpassed the state-of-the-art inverse halftoning methods based on deep learning. This indicates that the proposed image decomposition model is effective in obtaining high-quality continuous-tone images from halftone images. The proposed base layer design based on the GCM enables residual learning by narrowing the output brightness range. The structure-aware residual deblurring strategy can remove the blurring of the predicted base layer and restore the image structures effectively. The proposed SALDL is a new PRL for inverse halftoning. By contrast, the PSNR and SSIM of the DDN and PRL were lower than those of the proposed method. This confirmed that the DDN and PRL were restricted in terms of restoring the original images from low-quality base layers. Table II shows that the average PSNR of the U-net was slightly better than that of the PRL. This implies that the U-net is an extremely effective model for inverse halftoning. In other words, decomposing input halftoned images into multiple resolutions is an extremely effective approach. If the SRDAS and GCM-based residual subnetwork are built similarly as U-net, then the performance of the proposed method may be improved.

## IV. CONCLUSION

A new SALDL method for inverse halftoning was proposed. First, a new residual learning method based on the Gaussian convolution model was introduced for base layer generation. Compared with the additive model, which has been used for image denoising and rain removal, this Gaussian convolution model utilizes a statistical distribution, in which the image difference between the blurred original image and blurred halftone image with a Gaussian filter can possess a narrow brightness range. Second, a structure-aware residual deblurring strategy was presented. To remove the Gaussian blurring of the base layer and recover the image structures effectively, an image structure map predictor was designed to estimate the image structures from halftone patterns. This image structure map predictor enabled the entire network to be trained adaptively to local image structures; hence, noisy dot patterns on the flat areas were suppressed and local image structures such as lines and texts were described precisely. The experimental results confirmed that the proposed method surpassed state-of-the-art inverse halftoning methods based on deep learning, such as U-net, DCNN, DDN, and PRL. In addition, it was verified that the proposed image decomposition model was extremely effective in obtaining high-quality continuous-tone images from input halftone images.

REFERENCES

[1] L. Donghui, K. Takuma, H. Takahiko, T. Midori, and S. Kaku, "Texture-aware error diffusion algorithm for multi-level digital halftoning," *Journal of Imaging Science and Technology*, vol. 64, no. 5, pp. 50410-1-50410-9(9), Sept. 2020.

[2] Y. Wang, H. Huang, C. Wang, T. He, J. Wang, and M. H. Nguyen, "GIF2Video: Color dequantization and temporal interpolation of GIF images," *arXiv:1901.02840* [cs.CV], Jan. 2019.

[3] H. C. Do, B. G. Cho, S. I. Chien, and H. S. Tae, "Improvement of low gray-level linearity using perceived luminance of human visual system in PDP-TV," *IEEE Transactions on Consumer Electronics*, vol. 51, no. 1, pp. 204–209, Feb. 2005.

[4] W.-C. Kao, C.-H Liu, S.-C. Liou, J-C Tsai, and G.-H. Hou, "Towards video display on electronic papers," *Journal of Display Technology*, vol. 12, no. 2, Feb. 2016.

[5] C.-H. Son and H. Choo, "Watermark detection of clustered halftoned images via learned dictionary," *Signal Processing*, vol. 102, pp. 77-84, Sept. 2014.

[6] D. J. Lieberman and J. P. Allebach, "A dual interpretation for direct binary search and its implications for tone reproduction and texture quality," *IEEE Transactions on Image Processing*, vol. 9, no. 11, pp. 1950-1963, 2000.

[7] C. H. Son, "Inverse halftoning based on sparse representation," *Optics Letters*, vol. 37, no. 12, pp. 2352-2354, June 15, 2012.

[8] J.-M. Guo, H. Prasetyo, K. Wong, "Halftoning-based block truncation coding image restoration," *Journal of Visual Communication and Image Representation*, vol. 35, pp. 193-197, Feb. 2016.

[9] J.-M. Guo, "Watermarking in dithered halftone images with embeddable cells selection and inverse halftoning," *Signal Processing*, vol. 88, no. 6, pp. 1496–1510, June 2008.

[10] C.-H. Son, K. Lee, and H. Choo, "Inverse color to black-and-white halftone conversion via dictionary learning and color mapping," *Information Sciences*, vol. 299, pp. 1-19, Arp. 2015.

[11] J. Kopf and D, Lischinski, "Digital reconstruction of halftoned color comics," *ACM Transactions on Graphics*, vol. 31, no. 6, article no. 140, Nov. 2012.

[12] T. Remez, O. Litany, and A. Bronstein, "A picture is worth a billion bits: real-time image reconstruction from dense binary threshold pixels," in Proc. *IEEE International Conference on Computational Photography*, Evanston, USA, pp. 1-9, May 2016.

[13] M. Mese and P. P. Vaidyanathan, "Look-up table (LUT) method for inverse halftoning," *IEEE Transactions on Image Processing*, vol. 10, no. 10, pp. 1566–1578, Oct. 2001.

[14] T. D. Kite, N. Damera-Venkata, B. L. Evans, and A. C. Bovik, "A fast, high-quality inverse halftoning algorithm for error diffused halftones," *IEEE Transactions on Image Processing*, vol. 9, no. 9, pp. 1583–1592, Sep. 2000.

[15] R. Stevenson, "Inverse halftoning via MAP estimation," *IEEE Transactions on Image Processing*, vol. 6, no. 4, pp. 574–583, Apr. 1997.

[16] A. Foi, V. Katkovnik, K. Egiazarian, and J. Astola, "Inverse halftoning based on the anisotropic LPA-ICI deconvolution," in Proc. *Int. TICSP Workshop Spectral Meth. Multirate Signal Processing*, Vienna, Austria, pp. 49-56, 2004.

[17] C.-H Son and H. Choo, "Iterative inverse halftoning based on texture-enhancing deconvolution and error-compensating feedback," *Signal Processing*, vol. 93, no. 5, pp. 1126-1140, May 2013.

[18] P. G. Freitasa, Mylène C. Q. Fariasb, and Aletéia P. F. Araújo, "Enhancing inverse halftoning via coupled dictionary training," *Signal Processing: Image Communication*, vol. 49, pp. 1-8, Nov. 2016.

[19] C.-H Son and H. Choo, "Local learned dictionaries optimized to edge orientation for inverse halftoning," *IEEE Transactions on Image Processing*, vol. 23, no. 6, pp. 2542 – 2556, June 2014.

[20] Y. Zhang, E. Zhang, W. Chen, Y. Chen, J. Duan, "Sparsity-based inverse halftoning via semi-coupled multi-dictionary learning and structural clustering," *Engineering Applications of Artificial Intelligence*, vol. 72, pp. 43-53, June 2018.

[21] F. P. Jimenez, M. N. Miyatake, K. T. Medina, G. S. Perez, and H. P. Meana, "An inverse halftoning algorithms based on neural networks and atomic functions," *IEEE Latin America Transactions*, vol. 15, no. 3, pp. 488-495, March 2017.

[22] X. Hou and G. Qiu, "Image companding and inverse halftoning using deep convolutional neural networks," *arXiv:1707.00116* [cs.CV], July 2017.

[23] M. Xia and T.-T. Wong, "Deep inverse halftoning via progressively residual learning," in Proc. *Asian Conference on Computer Vision*, Australia, pp. 523-539, Dec. 2018.

[24] C.-H. Son, "Inverse halftoning through structure-aware deep convolutional neural networks," *Signal Processing*, vol. 173, pp. 1-7, Aug. 2020.

[25] J. Yuan, C. Pan, Y. Zheng, X. Zhu, Z. Qin, and Y. Xiao, "Gradient-guided residual learning for inverse halftoning and image expanding," *IEEE Access*, vol. 8, pp. 50995-51007, March 2020.

[26] L.-W. Kang, C.-W. Lin, and Y.-H. Fu, "Automatic single-image-based rain steaks removal via image decomposition," *IEEE Transactions on Image Processing*, vo. 21, no. 4, pp. 1742-1755, Apr. 2012.

[27] J. Lim, M. Heo, C. Lee, and C.-S. Kim, "Contrast enhancement of noisy low-light images based on structure-texture-noise decomposition," *Journal of Visual Communication and Image Representation*, vol. 45, pp. 107-121, May 2017.

[28] C.-H. Son and X.-P. Zhang, "Layer-based approach for image pair fusion," *IEEE Transactions on Image Processing*, vol. 25, no. 6, pp. 2866-2881, April 2016.

[29] J.-L. Starck, J. Fadili, and F. Murtagh, " The undecimated wavelet decomposition and its reconstruction," *IEEE Transactions on Image Processing*, vol. 16, no. 2, Feb. 2007.

[30] S. Paris, S. W. Hasinoff, and J. Kautz, "Local laplacian filters: edge-aware image processing with a Laplacian pyramid," *Communications of the ACM*, vol. 53, no. 3, pp 81–91, Feb. 2015.

[31] J. L. Stark, M. Elad, and D. L. Donoho, "Image decomposition via the combination of sparse representation and a variational approach," *IEEE Transactions on Image Processing*, vol. 14, no. 11, pp. 2675-2681, Oct. 2005.

[32] Y. Li, R. T. Tan, X. Guo, J. Lu and M. S. Brown, "Single image rain steak decomposition using layer priors", *IEEE Transactions on Image Processing*, vol. 26, no. 8, pp. 3874-3885, Aug. 2017.

[33] C. Tomasi and R. Manduchi, Bilateral filtering for gray and color Images, in Proc. *IEEE International Conference on Computer Vision*, 1998, pp. 839-846.

[34] S. Li, X. Kang, and J. Hu, "Image fusion with guided filtering," *IEEE Transactions on Image Processing*, vol. 27, no. 7, pp. 2864-2875, Jul. 2013.

[35] O. Ronneberger, P. Fischer, and T. Brox, "U-Net: Convolutional networks for biomedical image segmentation," *arXiv:1505.04597* [cs.CV] (2015).

[36] W. Lai, J. Huang, N. Ahuja and M. Yang, "Deep laplacian pyramid networks for fast and accurate super-resolution, in Proc. *IEEE Conference on Computer Vision and Pattern Recognition*, 2017, pp. 624-632.

[37] K. Zhang, W. Zuo, Y. Chen, D. Meng, and L. Zhang, "Beyond a gaussian denoiser: residual learning of deep CNN for image denoising," *IEEE Transactions on Image Processing*, vol. 26, no. 7, pp. 3142 – 3155, July 2017.

[38] X. Fu, J. Huang. D. Zeng, Y. Huang, X. Ding, J. Paisley, "Removing rain from single images via a deep detail network," in Proc. *IEEE International Conference on Computer Vision and Pattern Recognition*, Honolulu, USA, July 2016, pp. 1-9.

[39] M. Hradiš, J. Kotera, P. Zemčík and F. Šroubek, "Convolutional neural networks for direct text deblurring," In Proc. *British Machine Vision Conference*, Swansea, UK, Sept. 2015, pp. 6.1-6.13.

arXiv:2012.13894 [eess.IV]

[40] A. Vedaldi and K. Lenc, "Matconvnet: Convolutional neural networks for matlab," in Proc. *ACM international conference on Multimedia*, Brisbane Australia, pp. 689-692, Oct. 2015.
[41] Z. Wang, A. C. Bovik, H. R. Sheikh, and E. P. Simoncelli, "Image quality assessment: from error visibility to structure similarity," *IEEE Transactions on Image Processing*, vol. 13, no. 4, pp. 600–612, Apr. 2004.
[42] J.-H. Kwon, C.-H. Son, Y.-H. Cho and Y.-H. Ha, "Text-enhanced error diffusion using multiplicative parameters and error scaling factor," *Journal of Imaging Science and Technology*, vol. 50, no. 5, pp. 437–447, Sep.-Oct. 2006.
[43] S. Ruder, "An overview of gradient descent optimization algorithms" *arXiv:1609.04747* [cs.LG], Sept. 2016.